\documentclass[twocolumn]{revtex4-1}
\usepackage[utf8]{inputenc}
\usepackage{graphicx}
\usepackage{amssymb}
\usepackage{amsmath}
\usepackage{verbatim}
\usepackage{ulem}
\begin{document}

\title {
Unfolding of antiferromagnetic phases and  multicritical points in a two-orbital model for Uranium compounds under pressure and magnetic field}
\author{S. G. Magalhaes}
\affiliation{Instituto de Física, Universidade Federal do Rio Grande do Sul, 91501-970
Porto Alegre, RS, Brazil}
\author{A. C. Lausmann and E. J. Calegari}
\affiliation{Departamento de F\'{\i}sica - UFSM, 97105-900, Santa Maria, RS, Brazil}%
\author{P. S. Riseborough}%
\affiliation{Physics Department, Temple University, Philadelphia, Pennsylvania 19122, USA}
\date{\today}

\begin{abstract}

We investigate the occurrence of multicritical points under pressure and magnetic field
in a model that describes two 5f bands (of either $\alpha$ or $\beta$ characters) which hybridize with a single itinerant conduction band. The 5f-electrons interact through Coulomb and exchange terms. The AF order parameter is a N\'eel vector, which is assumed to be fixed by an Ising anisotropy. The applied magnetic field is transverse to the anisotropy axis. Without field, our results for the  temperature - pressure phase diagram show that, at low temperatures, a first-order phase transition occurs between two distinct antiferromagnetic phases, AF$_1$ and AF$_2$, as the pressure is increased. The two phases are characterized by the gaps of bands $\alpha$ and $\beta$ given by  $\Delta_{\alpha}$ and $\Delta_{\beta}$, respectively. The AF$_1$ phase occurs when $\Delta_{\beta}>\Delta_{\alpha}>0$, while in the AF$_2$ phase, the gaps satisfy $\Delta_{\alpha}>\Delta_{\beta}>0$. The application of a magnetic field produces a drastic change in the phase diagram. The  AF1 and AF2 phases separate with the latter acquiring a dome shape which is eventually suppressed for large values of the applied field. The evolution of the phase diagram under pressure, without and with magnetic field, shows the presence
of multicritical points. Our results show that the evolution of these multicritical points by the simultaneous application of pressure and field is also drastic with the suppression of some multicritical points and the emergence of others ones. We believe that these results may have relevance for the growing field of multicritical points (classical and quantum) in the physics of Uranium compounds.

\end{abstract}



\maketitle

\section{Introduction}

The uniqueness of the $5f$-electrons physics is due to the dual localized-delocalized character \cite{Moore2009}.
Uranium compounds display a variety of quantum states of matter, such as magnetism (localized and itinerant)  \cite{Santini1999}, superconductivity \cite{Pfleiderer2009}
and the enigmatic Hidden Order\cite{Mydosh2011,Oppeneer2014}. These phases can be tuned by pressure (hydrostatic or chemical) and magnetic field.
The plethora of distinct types of ordering hosted by uranium compounds makes these systems a natural ground for the appearance of multicritical points. Likewise, the presence of a specific kind of multicritical point could be useful in the clarification of an unconventional symmetry breaking that may exist in
uranium compounds. There is a possibility  that the multicritical points driven by thermal fluctuations might evolve into quantum multicritical points \cite{Belitz2017}
as pressure or magnetic field increase \cite{Kotegawa2011} leading to deviations of Fermi liquid behaviour \cite{Gegenwart2008,Misawa2008}.

The presence of multiple $5f$ orbitals in uranium compounds has important consequences for their physics \cite{Nozieres1980,Cox1987}.
The Under-screened Anderson lattice model (UALM) we use consists of two degenerate narrow $5f$ bands (denoted by $\chi=\alpha,\beta$), which acquire itinerant character by direct hopping between neighboring $5f$ shells. The resulting two 
bands are also hybridized with a single itinerant conduction band. The interaction is composed of the Coulomb interaction between electrons in the same $5f$ bands and the Hund's rule exchange interaction between electrons in distinct ones. 
 This  model 
considering the ground state being a triplet (S=1) can be related via a Schrieffer-Wolff transformation with the Underscreened Kondo Lattice 
\cite{Thomas2011} which has 
successfully described the coexistence of the Kondo effect and  ferromagnetism found in uranium monochalcogenides \cite{Perkins2007}.  
 Most important for the purposes of this work that does not deal with the Kondo effect is that
the UALM can describe, not only the antiferromagnetic ordering observed in the uranium-pnictides \cite{Giannakis2019,Joyce2004,Gukasov1999} and UIrSi$_3$ \cite{Honda2018}, but has also been proposed to describe the Hidden Order phase of the URu$_2$Si$_2$ \cite{Riseborough2012}.

 Indeed, the UALM is 
suited for the investigation of time-reversal symmetry breaking as source of unfolding of phases and multicritical points. The  Hund's rule exchange interaction term is essential to make the model spin-rotationally invariant \cite{Caroli1969} and opens distinct routes to long-range ordering.
As an example, a phase transition can be driven by the spin-flip part of Hund's rule exchange interaction, breaking spin-rotational and space-translational symmetries but preserving the time-reversal symmetry. As a result, a novel ordered state can be stabilized in which there is spontaneous $5f$ inter-band mixing, that does not involve magnetic order. This novel type of long-ranged order has been proposed as describing the Hidden Order phase in URu$_2$Si$_2$ \cite{Riseborough2012}. The interaction terms can also produce conventional antiferromagnetic long-range order in the UALM. The here called intra-orbital antiferromagnetic phases (IOAF) appear below a magnetic phase transition at which spin-rotational, space-translational and time-reversal symmetries are broken. This transition gives rise to not one but two distinct competing IOAF phases which have spin gaps at the same ordering wave vector. Therefore, in the transition between the two IOAF phases, no further symmetries are broken.

In this work, we investigate the temperature - pressure - magnetic field phase diagram of the IOAF phases within a mean-field approximation. 
The objective is 
to explore 
the UALM 
to describe the unfolding of an itinerant antiferromagnetic phase and the subsequent competition between the two unfolded different antiferromagnetic phases.
In particular, We focus on multicritical points which can appear from that competition and how they evolve with pressure and field. 
The symmetry contained in the model that can be broken
(as the 
time reversal one)
shall be reflected in the parity properties of the order parameters which condition the emergence of multicritical points.
We stress  that the subject of multicritical points can be 
 connected to real Uranium compounds.
For instance, a bicritical point was found in URu$_2$Si2 \cite{Mydosh2011} when pressure is applied and tricritical (TCP) points were found in USb$_2$  \cite{Stillwell2017}, UN \cite{Shrestha2017}, UAu$_2$Si$_2$ \cite{Valiska2018} and URu$_2$Si$_2$ \cite{Correa2012} when field is applied. However,  the simultaneous effect of pressure and field on multicritical points has received little theoretical attention (as an exception, see Ref. \cite{Haule2010}) in the physics of Uranium compounds.

We assume that the bandwidth $W$ can be varied by the application of pressure while the hybridization, the Coulomb and the Hund's rule interactions remain constant. We also make the following assumptions: (i) The hybridization matrix elements are $k$-independent. As a consequence, one may transform the basis of the $5f$ states into a new basis in which a linear combination of f orbitals hybridize and the remaining orthogonal states do not. 
The asymmetric hybridization breaks the symmetry between the $5f$ bands, so intra-band nesting may occur simultaneously for both bands but, when $W$ increases, one band may become depart from the perfect nesting condition and, hence have a reduced moment. Ultimately, above some value of $W$, both bands might not satisfy the nesting condition and the material might become non-magnetic.
(ii) the IOAF has a N\'eel order parameter which is fixed by an Ising-like anisotropy. This assumption introduces a magnetic anisotropy which, in fact, is observed in some uranium compounds \cite{Schoenes1981,Havela1992,Maskova2019}. As a consequence, there are two types of field effects in the IOAF 
\cite{Calegari2017,Rise2014}. For a field aligned with the easy axis, the Zeeman splitting between the spin-up and spin-down IOAF sub-bands increases as the field increases. On the other hand, for a field along a perpendicular
direction, there is  a spin-dependent momentum-shift of the IOAF bands. In both cases, the nesting condition may no longer be satisfied.  We chose the last possibility due to its spin-flipping effects.

In the 
case of two distinct IOAF (denoted as AF$_1$ and AF$_2$), 
since we are dealing with two 
bands,
the phase transition AF$_1$ $\rightarrow$ AF$_2$
would necessarily imply that the two spins gaps abruptly interchange their sizes. 
Eventually, as pointed out above, a further variation of $W$ can cause the complete suppression of the IOAF ordering. The sequence of transitions AF$_1$ $\rightarrow$ AF$_2$  $\rightarrow$ PM should involve changes in the structure of the AF bands and, therefore, should be accompanied by Fermi Surface (FS) reconstruction. One may also expect that other sequences of phase transition involving IOAF caused by increasing the field $h_x$ are also related to changes in the electronic structure.

We remark that the phase diagram  temperature vs pressure of the IOAF, in the absence of a magnetic field, can be deduced from general arguments based on a Landau free energy with two order parameters.
In the simplest situation, the two AF order parameters are linearly coupled \cite{Shah2000}, since both break time-reversal symmetry. In contrast, two  order parameters which are even under time-reversal couple quadratically leading to bicritical or tetracritical points \cite{Chaikin1995}. In the first case, at very low temperatures, there should be a line of first-order transitions separating two distinct AF phases as an intensive thermodynamic parameter $g$ is varied. The line of first-order transitions
ends at a critical end point (CEP) located at ($T_{CEP}$, $g_{CEP}$). Actually, this kind of scenario was proposed by Mineev and Zhitomirsky (see Ref. \cite{Mineev2005})
to account for the temperature-pressure phase diagram of the URu$_2$Si$_2$ \cite{Mydosh2011}. In their description, the experimentally determined phase transition between the Hidden Ordered and the antiferromagnetic phases involves two types of antiferromagnetism,  itinerant antiferromagnetism with small magnetic moments and localized antiferromagnetism with large magnetic moments. Unfortunately, there is strong evidence that the Hidden Order is not explainable in terms of conventional antiferromagnetic phases.

This paper is organized as follows: the UALM is presented in the section \ref{2}. In section \ref{3}, we derive the Green's functions and the free energy. Next, in the section \ref{4}, we obtain the order parameters and gaps for zero and non-zero values of the transverse field $h_x$. A discussion of the numerical results follows in section \ref{5}. The conclusions and other remarks are found in section \ref{6}.

\section{Model}\label{2}

We shall investigate a generic form of the UALM which is written as
\begin{equation}
\hat{H} \ = \ \hat{H}_f \ + \ \hat{H}_d \ + \ \hat{H}_{fd}.
\label{hamilt}
\end{equation}
The $f$ electron part of Hamiltonian $\hat{H}_f$ is given by
$\hat{H}_f= \hat{H}_{f,0}+ \hat{H}_{f,int}$,
where the non-interacting part $H_{0,f}$ describes two degenerate narrow $f$ bands and is expressed as
\begin{eqnarray}
\hat{H}_{f,0} = \sum_{\underline{k},\sigma}\sum_{\chi} \ E_{f}^{\chi}(\underline{k}) \
f^{\dag \chi}_{\underline{k},\sigma}f^{\chi}_{\underline{k},\sigma}.
\label{eq0}
\end{eqnarray}
The $\chi$-bands ($\chi=\alpha$ and $\beta$) in Eq. (\ref{eq0}) obey the intraband and interband nesting
condition
$E^{\chi}_{f}(\underline{k}+\underline{Q})=-E^{\chi'}_{f}$(\underline{k})
where $\chi=\chi'$ (intraband) or $\chi\neq\chi'$ (interband). The vector $\underline{Q}$ is a commensurate momentum transfer
in the Brillouin zone. The interaction between the $f$-electrons is described by
\begin{eqnarray}
\hat{H}_{f, int} \ = \ \bigg(\frac{U - J}{2 N}\bigg) \
\sum_{\underline{k},\underline{k}',\underline{q},\sigma,\chi\ne\chi'}
\ f^{\dag,\chi}_{\underline{k}+\underline{q},\sigma} \ f^{\chi}_{\underline{k},\sigma} \
f^{\dag,\chi'}_{\underline{k}'- \underline{q},\sigma} f^{\chi'}_{\underline{k}',\sigma}
\nonumber \\
+ \ \bigg({ U \over 2 N}\bigg)
\sum_{\underline{k},\underline{k}',\underline{q},\sigma,\chi,\chi'}
\ f^{\dag,\chi}_{\underline{k}+\underline{q},\sigma} \ f^{\chi}_{\underline{k},\sigma} \
f^{\dag,\chi'}_{\underline{k}'-\underline{q},-\sigma} f^{\chi'}_{\underline{k}',-\sigma}
\nonumber \\ +  \bigg({J \over 2 N}\bigg)
\sum_{\underline{k},\underline{k}',\underline{q},\sigma,\chi\ne\chi'} \
f^{\dag,\chi}_{\underline{k}+\underline{q},\sigma} \  f^{\chi'}_{\underline{k},\sigma} \
f^{\dag,\chi'}_{\underline{k}'-\underline{q},-\sigma} \ f^{\chi}_{\underline{k}',-\sigma}.\nonumber\\
\label{Hfintreal}
\end{eqnarray}where $U$ is the screened Coulomb interaction and $J$ is the Hund's rule exchange. The conduction electron Hamiltonian $\hat{H}_d$ is expressed as
\begin{equation}
    \hat{H}_d \ = \ \sum_{\underline{k},\sigma} \ \epsilon_d(\underline{k}) \ d^{\dag}_{\underline{k},\sigma} \ d_{\underline{k},\sigma}
\label{Hd}
\end{equation}
where $\epsilon(\underline{k})$ describes the dispersion relation of conduction electrons labeled by the Bloch wave vector $\underline{k}$.
The last term in Eq. (\ref{hamilt}) describes the on-site hybridization process in the UALM model by
\begin{equation}
    \hat{H}_{fd} \ = \ \sum_{\underline{k},\sigma}\sum_{\chi=\alpha\beta} \ \bigg( \
V_{\chi}(\underline{k})
\ f^{\dag,\chi}_{\underline{k},\sigma} \ d_{\underline{k},\sigma} \ + \
V^*_{\chi}(\underline{k}) \ d^{\dag}_{k,\sigma} \ f^{\chi}_{\underline{k},\sigma} \ \bigg).
\label{Hhyb}
\end{equation}\\

In the present work, both the 5f band $E_f^{\chi}(\underline{k})=\tilde{\epsilon}_{f}+\epsilon_f(\underline{k})$ and the conduction one $\epsilon_d(\underline{k})$  refer to a cubic lattice while $\tilde{\epsilon}_{f}$ is the band center. Thus
\begin{equation}
\epsilon_{b}(\underline{k})=-2t_b[\cos(k_xa)+\cos(k_ya)+\cos(k_za)]
\label{Ek}
\end{equation}
in which $b=f$ or $d$, and $a$ is the lattice parameter.

The simplest possibility of a IOAF ordering with Ising anisotropy in a cubic lattice can be introduced by assuming that the lattice is bipartite. Therefore, we consider that
\begin{equation}
n^{\chi}_{f,\underline{q},\sigma}=\frac{n^{\chi}_{f}}{2} \ \delta_{\underline{q},0} +
m^{\chi}_{f} \ \eta(\sigma) \ \delta_{\underline{q},\pm \underline{Q}}
\label{nq}
\end{equation}
where $n^{\chi}_{f}=n^{\chi}_{f,\uparrow}+n^{\chi}_{f,\downarrow}$  ($n^{\chi}_{f}$ is the $f$-electron average occupation of the $\chi$-band), $\eta(\uparrow)=+1$ or $\eta(\downarrow)=-1$ and $\underline{Q}=(\pi/a,\pi/a,\pi/a)$, is a commensurate nesting vector. Therefore, the  modulation of the expectation value of the $z$-component f-electron spin density operator in real space for each orbital is $\langle\hat{S}^{\chi}_{z,\underline{r}_j}\rangle=m^{\chi}_{f} \ e^{i\underline{Q}\ . \
\underline{r}_{j}}$. The IOAF order parameters, i.e., the staggered magnetizations $m_f^{\alpha}$ and $m_f^{\beta}$ are obtained from
\begin{equation}
m_f^{\chi}={1 \over 2}(n^{\chi}_{f,\underline{Q},\uparrow}-n^{\chi}_{f,\underline{Q},\downarrow}).
\label{PO}
\end{equation}

\section{ General Formulation}
\label{3}

We include an applied magnetic field oriented transverse to the $z$-axis, which introduces an additional term into the Hamiltonian
$\hat{H}_{ext}=\hat{H}_{ext}^{f}+\hat{H}_{ext}^{d}$ where
\begin{equation}
\hat{H}_{ext}^{f}=-\Gamma_f\sum_{\underline{k}} \ (f^{\dag}_{\underline{k},\uparrow} \ f_{\underline{k},\downarrow}\ + \ f^{\dag}_{\underline{k},\downarrow} \ f_{\underline{k},\uparrow})
\label{campo}
\end{equation}
with
\begin{equation}
\Gamma_f= g_{f} \mu_B h_x.
\label{Gammaf}
\end{equation}
The term $\hat{H}_{ext}^{d}$ is the same as Eq. (\ref{campo}), except that the $f$-operators and the gyromagnetic factor $g_f$ are replaced by $d$-operators and $g_d$, respectively.

The temporal and spatial Fourier transform of the single-electron f-f Green's function, within the Hartree-Fock approximation, satisfy the equations of motion
given by:
\begin{eqnarray}
[ \ \omega \ - \ \tilde{E}_{f}^{\alpha}(\underline{k}) \ ] \ G^{\alpha,\chi'}_{ff,\sigma}(\underline{k},\underline{k}',\omega) =
\delta^{\alpha,\chi'}  \ \delta_{\underline{k},\underline{k}'} \ \delta_{\sigma,\sigma'}\nonumber \\
+ \ V_{\alpha}(\underline{k})  \ G^{\chi'}_{df,\sigma\sigma'}(\underline{k},\underline{k}',\omega)
- \ \Gamma_f \ G^{\beta,\chi'}_{ff,-\sigma,\sigma}(\underline{k},\underline{k}',\omega)\nonumber \\
+ \phi^{\alpha}_{\sigma} \
G^{\alpha,\chi'}_{ff,\sigma,\sigma'}(\underline{k}+\underline{Q},\underline{k}',\omega)\nonumber\\
\label{Galpha}
\end{eqnarray}
and
\begin{eqnarray}
[ \ \omega \ - \ \tilde{E}_{f}^{\beta}(\underline{k}) \ ] \ G^{\beta,\chi'}_{ff,\sigma,\sigma'}(\underline{k},\underline{k}',\omega)=
  \delta^{\beta,\chi'} \ \delta_{\underline{k},\underline{k}'} \ \delta_{\sigma,\sigma'}\nonumber \\
+ \ V_{\beta}(\underline{k}) \ G^{\chi'}_{df,\sigma,\sigma'}(\underline{k},\underline{k}',\omega)
- \ \Gamma_f \ G^{\alpha,\chi'}_{ff,-\sigma\sigma'}(\underline{k},\underline{k}',\omega) \nonumber \\
+ \phi^{\beta}_{\sigma} \ G^{\beta,\chi'}_{ff,\sigma,\sigma'}(\underline{k}+\underline{Q},\underline{k}',\omega).\nonumber\\
\label{Gbeta}
\end{eqnarray}
The spin-independent Hartree-Fock dispersion relation $\tilde{E}_{f}^{\chi}(\underline{k})$ is given by
\begin{equation}
    \tilde{E}_{f}^{\chi}(\underline{k}) \ = \ E_f^{\chi}(\underline{k}) \ + \
\sum_{\chi'}  \bigg( (U - J)  \frac{n^{\chi'}_{f}}{2}
    (1-\delta^{\chi,\chi'})  +  U  \frac{n^{\chi'}_{f}}{2} \bigg)
\label{Erenorm}
\end{equation}
where the real function $\phi^{\chi}_{\sigma}$ is given by
\begin{equation}
\phi^{\chi}_{\sigma} \ = \ \sum_{\chi'}  \bigg( U  m^{\chi'}  \eta(-\sigma) +  (
U  -  J )   m^{\chi}  (1-\delta^{\chi,\chi'})  \eta(\sigma)\bigg).
\label{GapNeel}
\end{equation}

The mixed $f-d$ Green's function satisfies the following equation
\begin{eqnarray}
&&     [ \ \omega \ - \ \epsilon(\underline{k}) \ ] \ G^{\chi'}_{df,\sigma,\sigma'}(\underline{k},\underline{k}',\omega) \ =
     \ V_{\alpha}(\underline{k})^* \ G^{\alpha,\chi'}_{ff,\sigma,\sigma'}(\underline{k},\underline{k}',\omega) \nonumber\\
&&    + \ V_{\beta}(\underline{k})^* \ G^{\beta,\chi'}_{ff,\sigma,\sigma'}(\underline{k},\underline{k}',\omega)
-\Gamma_d \ G^{\chi'}_{df,-\sigma,\sigma'}(\underline{k},\underline{k}',\omega).
\label{Gmix}
\end{eqnarray}

We will choose a basis set for the $f$ orbitals, such that $V_{\beta}({\underline{k}})=0$ and $V_{\alpha}({\underline{k}})=V_{\alpha}$
simply to avoid the transformation to a new basis set. The choice of basis states should not change the main physical results, as discussed in ref. \cite{Riseborough2012}.
The Green's function equation of motions given in Eqs. (\ref{Galpha})-(\ref{Gmix}) form a closed set of equations, which can be solved exactly. The equations can be expressed in the matrix form
\begin{equation}
\underline{\underline{\varPi}}^{\chi}(\underline{k},\omega) \ \underline{\underline{G}}^{\chi\chi'}(\underline{k},\underline{k'},\omega)=\underline{\underline{\delta}}^{\chi'}(\underline{k},\underline{k'})
\label{matrix}
\end{equation}
where
\[ \underline{\underline{G}}^{\chi\chi'}(\underline{k},\underline{k'},\omega)=\left( \begin{array}{c}
G_{ff,\sigma\sigma'}^{\chi\chi'}(\underline{k},\underline{k'},\omega) \\
G_{ff,\sigma\sigma'}^{\chi\chi'}(\underline{k}+\underline{Q},\underline{k'},\omega)\\
G_{ff,-\sigma\sigma'}^{\chi\chi'}(\underline{k},\underline{k'},\omega) \\
G_{ff,-\sigma\sigma'}^{\chi\chi'}(\underline{k}+\underline{Q},\underline{k'},\omega)
\end{array} \right) \]
\begin{widetext}
\[ \underline{\underline{\varPi}}^{\chi}(\underline{k},\omega)=\left( \begin{array}{cccc} \omega-\tilde{E}_{f}^{\chi}(\underline{k})- \xi_{\Gamma}^{'\chi}(\underline{k}) &- \phi^{\chi}_{\sigma} & \gamma_{\Gamma}^{'\chi}(\underline{k}) & 0\\
- \phi^{\chi}_{\sigma} & \omega -\tilde{E}_{f}^{\chi}(\underline{k}+\underline{Q})-\xi_{\Gamma}^{'\chi}(\underline{k}+\underline{Q})
& 0 & \gamma_{\Gamma}^{'\chi}(\underline{k}+\underline{Q}) \\
\gamma_{\Gamma}^{'\chi}(\underline{k}) & 0 & \omega- \tilde{E}_{f}^{\chi}(\underline{k})-\xi_{\Gamma}^{'\chi}(\underline{k}) & -\phi^{\chi}_{-\sigma} \\
0 & \gamma_{\Gamma}^{'\chi}(\underline{k}+\underline{Q}) & -\phi^{\chi}_{-\sigma}
& \omega -\tilde{E}_{f}^{\chi}(\underline{k}+\underline{Q})-\xi_{\Gamma}^{'\chi}(\underline{k}+\underline{Q})  \end{array} \right) \]
\end{widetext}

\[\underline{\underline{\delta}}^{\chi'}(\underline{k},\underline{k'}) =\left( \begin{array}{c}
\delta^{\chi\chi'}\delta_{\underline{k},\underline{k'}}\delta_{\sigma\sigma'} \\
\delta^{\chi\chi'}\delta_{\underline{k}+\underline{Q},\underline{k'}}\delta_{\sigma\sigma'}\\
\delta^{\chi\chi'}\delta_{\underline{k},\underline{k'}}\delta_{-\sigma\sigma'} \\
\delta^{\chi\chi'}\delta_{\underline{k}+\underline{Q},\underline{k'}} \delta_{-\sigma\sigma'}
\end{array} \right) \]
 with $\xi_{\Gamma}^{'\chi}(\underline{k})= \xi_{\Gamma}^{\chi}(\underline{k})(\delta_{\chi\alpha}+(1-\delta_{\chi\beta}))$, $\gamma_{\Gamma}^{'\chi}(\underline{k})=\gamma_{\Gamma}^{\chi}(\underline{k})(\delta_{\chi\alpha}+(1-\delta_{\chi\beta}))$ where
\begin{equation}
\xi_{\Gamma}^{\chi}(\underline{k})= \frac{(\omega-\epsilon_d(\underline{k})) \ |V_{\chi}|^2}{(\omega-\epsilon_d(\underline{k}))^{2}-\Gamma_{d}^{2}}
\label{xi}
\end{equation}
and
\begin{equation}
\gamma_{\Gamma}^{\chi}(\underline{k})= \Gamma_{f}-\frac{\Gamma_{d} \  |V_{\chi}|^2}{(\omega-\epsilon_d(\underline{k}))^{2}-\Gamma_{d}^{2}}.
\label{gamma}
\end{equation}

\section {Intra-orbital Antiferromagnetic (IOAF) phase}
\label{4}

From now on, we will focus on the  IOAF phases and their associated phase transitions. 

\subsection{Order Parameters and Gaps
with $h_x=0$}

The IOAF order parameters follow directly from the correlation functions $n^{\chi}_{f,\underline{Q},\sigma}$ (see Eq. (\ref{PO})), which can be expressed as
\begin{eqnarray}
n^{\chi}_{f,\underline{Q},\sigma}=
\frac{1}{N} \sum_{\underline{k},\sigma}
 \oint \frac{d\omega}{2\pi i} f(\omega) \ G^{\chi\chi}_{ff,\sigma}(\underline{k},\underline{k}+\underline{Q},\omega).
\label{GF}
\end{eqnarray}
The contour of the path integral encircles the real axis without enclosing any poles  of the Fermi-Dirac distribution.

The correlation function $n^{\beta}_{f,\underline{Q},\sigma}$ is found from the Green's function given in Eq. (\ref{Gb}). Therefore,
from Eq. (\ref{PO}), one can obtain:
\begin{eqnarray}
m^{\beta}_{f}=  (U  m_{f}^{\beta} \ + \ J m_{f}^{\alpha}) \
\chi^{\beta\beta}_{f} (\underline{Q},0)
\label{mb}
\end{eqnarray}
where
\begin{eqnarray}
\chi^{\beta\beta}_{f}(\underline{Q},0)=
{1 \over N}\sum_{\underline{k}} \
{f(E^{-}(\underline{k}))-f(E^{+}(\underline{k})) \over
E^{+}(\underline{k}) \ - \ E^{-}(\underline{k}) }
\label{XQb}
\end{eqnarray}
and $f(\omega)$ is the Fermi function.

The staggered magnetization of the $\alpha$-bands can be derived in a similar manner to  Eq.(\ref{mb}). The result is
\begin{eqnarray}
m_{f}^{\alpha}=(U m_{f}^{\alpha} \ + \ J
m_{f}^{\beta})
\ \chi^{\alpha\alpha}_{f}(\underline{Q},0)
\label{ma}
\end{eqnarray}
where $\chi^{\alpha\alpha}_{f}(\underline{Q},0)$ is now given as
\begin{eqnarray}
& &\chi^{\alpha\alpha}_{f}(\underline{Q},0)=\nonumber \\
& &\frac{1}{2N} \sum_{\underline{k},\sigma}
 \oint \frac{d\omega}{2\pi i} f(\omega){(\omega-\epsilon_d(\underline{k}))(\omega-\epsilon_d(\underline{k}+\underline{Q})) \over
D^{\alpha}(\omega,\underline{k})}.\nonumber \\
\label{XQa}
\end{eqnarray}
with $D^{\alpha}$  given in Eq. (\ref{Da}).
The spin-independent quasiparticle bands are given by the solutions of $D^{\alpha}(\underline{k}, \omega)=0$.

Alternatively, one can formulate the self-consistency equations in terms of the gaps
$\Delta_{\alpha(\beta)}$ since
\begin{equation}
\phi^{\alpha(\beta)}_{\uparrow\downarrow}=
\ \mp \ \Delta_{\alpha(\beta)}
\label{phi}
\end{equation}
in which
\begin{eqnarray}
\Delta_{\alpha(\beta)}
=U m_f^{\alpha(\beta)}+J m_f^{\beta(\alpha)}.
\label{phi1}
\end{eqnarray}
The Hund's rule interaction couples the gap of a given band to the staggered magnetization of the other band. Using Eqs. (\ref{mb}), (\ref{XQb}), (\ref{ma}) and (\ref{XQa}), one may write the coupled equations for the gaps $\Delta_{\alpha}$ and $\Delta_{\beta}$ as
\begin{eqnarray}
U\Delta_{\alpha(\beta)}-J\Delta_{\beta(\alpha)} =
 (U^{2}-J^{2})
\Delta_{\alpha(\beta)}\chi^{\alpha\alpha(\beta\beta)}_{f}(\underline{Q},0).\nonumber \\
\label{GapNb}
\end{eqnarray}

 \subsection{Order Parameters and Gaps with
$h_x\neq 0$
}

For $h_x\neq 0$, the pole structure of the Green's functions is much more complex, and is shown in Appendix A. For finite fields, the Green's functions
$G^{\beta\beta}_{ff,\sigma}(\underline{k},\underline{k}+\underline{Q},\omega)$ and $G^{\alpha\alpha}_{ff,\sigma}(\underline{k},\underline{k}+\underline{Q},\omega)$
shown in Eqs. (\ref{A2}) and (\ref{A6}), can be used to obtain the order parameters $m_{\alpha}$, $m_{\beta}$ and the gaps
following the same steps outlined in Section (\ref{4})-A. We assume that the $d$-conduction electron band is uncorrelated and wider than the correlated $f$-bands. We note that the magnetic field on the d-electrons, $\Gamma_d$, affects the order parameters  $m^{\alpha}$ and $m^{\beta}$, mainly through the effect of the hybridization $V_{\alpha}$, and is small compared to the $\alpha$ and $\beta$ bandwidths. Therefore, it is reasonable to disregard the effects of $\Gamma_d$ on $m^{\alpha}$ and $m^{\beta}$.

\section{Free Energy }

In the Hartree-Fock approximation, the free energy can be expressed in terms of the gaps by:
\begin{equation}
f_{HF}= \Omega(T,\mu)+ \mu N_{tot} +\frac{N}{U^2-J^2} \  [U(\Delta_{\alpha}^2 + \Delta_{\beta}^2)-2J\Delta_{\alpha}\Delta_{\beta}]
    \label{freeenergy}
\end{equation}
where $\mu$ is the chemical potential,  $N_{tot}=n_f^{\alpha}+n_f^{\beta}+n_d$ ($n_d$ is the average occupation of the conduction electrons) and
\begin{equation}
 \Omega(T,\mu)= -k_{B} T \frac{1}{N}\sum_{\underline{k}}\sum_{\gamma} \ln[1+ e^{-\frac{(E^{\gamma}-\mu)}{k_{B} T}}],
     \label{freeenergy0}
\end{equation}
where $N$ is the number of sites in the lattice. The quasi-particles energies $E^{\gamma}$ are obtained from:
(i) for $h_x=0$, from Eq. (\ref{emm}) and the roots of $D^{\alpha}(\omega,\underline{k})=0$ (see Eq. (\ref{Da}));
(ii) for $h_x\neq 0$, from the Appendix \ref{appendixa}, as roots of $|A^{\beta}|=0$ (see Eq. (\ref{A4})) and $|A^{\alpha}|=0$ (see Eq. (\ref{A9})) within the approximations  $\gamma_{\Gamma}^{\chi}(\underline{k})\approx \Gamma_f$ and $\xi_{\Gamma}^{\chi}(\underline{k})\approx \frac{ |V_{\chi}|^2}{(\omega-\epsilon_d(\underline{k}))}$ where we note that $\Gamma_d \approx 0$. It should be remarked that both $|A^{\alpha(\beta)}|$ depend directly on $\Delta^{2}_{\alpha(\beta)}$ (see Eq. (\ref{phi})).

\section{Instability of the Paramagnetic Metallic (PM) phase}

In the general case ($h_x=0$ and $h_x\neq 0$), a second-order instability of the PM phase can be determined from the linearized equations for the order parameters. The instability occurs when
\begin{eqnarray}
(1-U\chi^{\alpha\alpha(0)}_{f}(\underline{Q},0))(1-U\chi^{\beta\beta(0)}_{f}(\underline{Q},0))\nonumber\\
=J^{2}\chi^{\alpha\alpha(0)}_{f}(\underline{Q},0)\chi^{\beta\beta(0)}_{f}(\underline{Q},0).
\label{StoSDW}
\end{eqnarray}
This equation determines the N\'eel temperature $T_N$.

\section {Results}
\label{5}

\begin{figure}[t]
\centering
\includegraphics[angle=0,width=7.5cm]{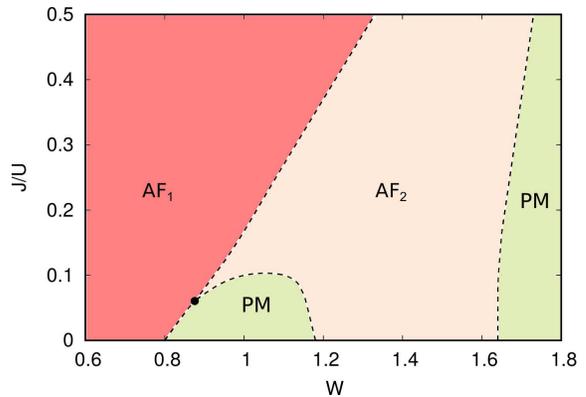}
\caption{The phase diagram for $J/U$ versus the band width $W$ and $T=0$. All the phases transition are first-order.
}
\label{fig01}
\end{figure}
\begin{figure}[h]
\centering
\includegraphics[angle=0,width=8.5cm]{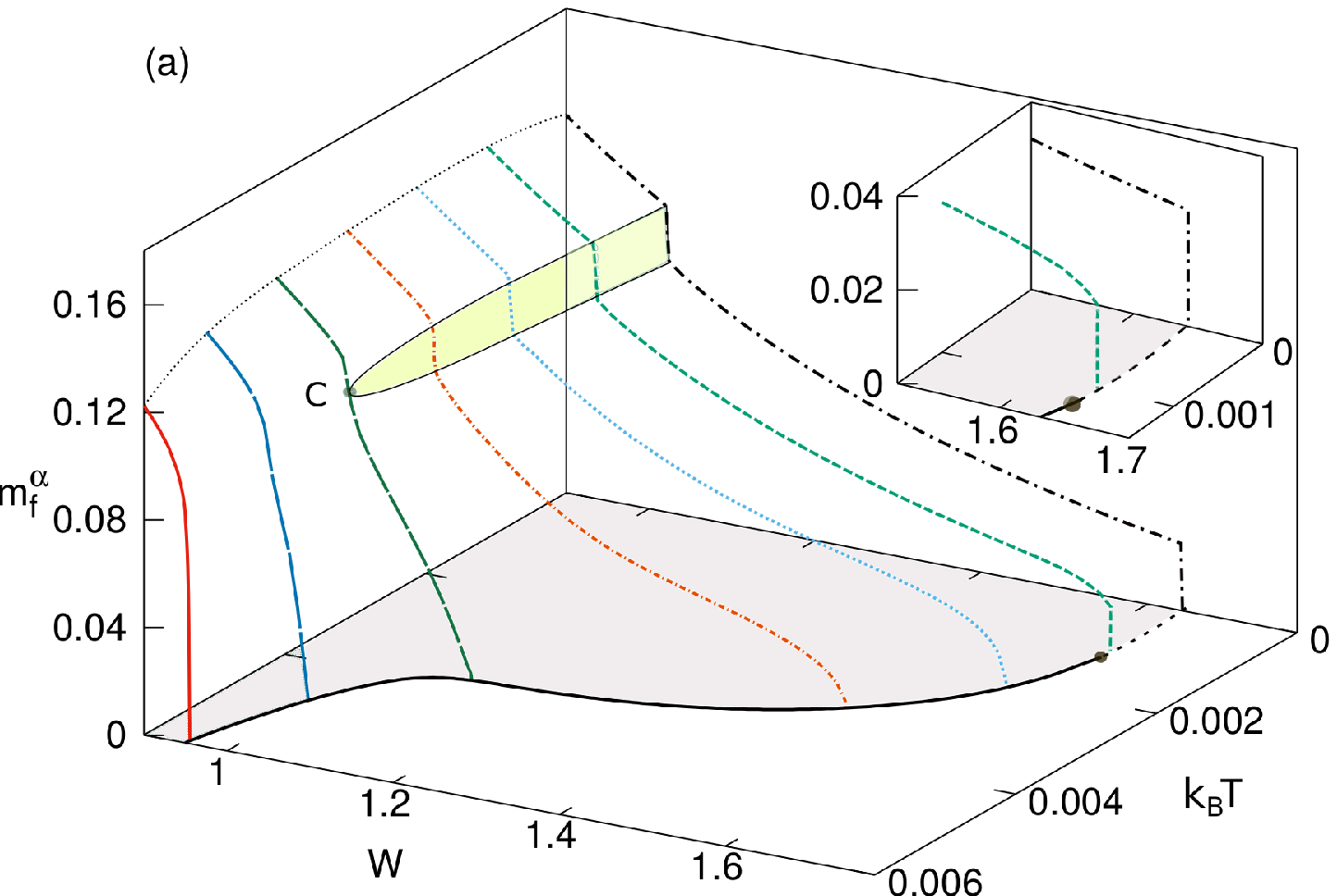}
\includegraphics[angle=0,width=8.5cm]{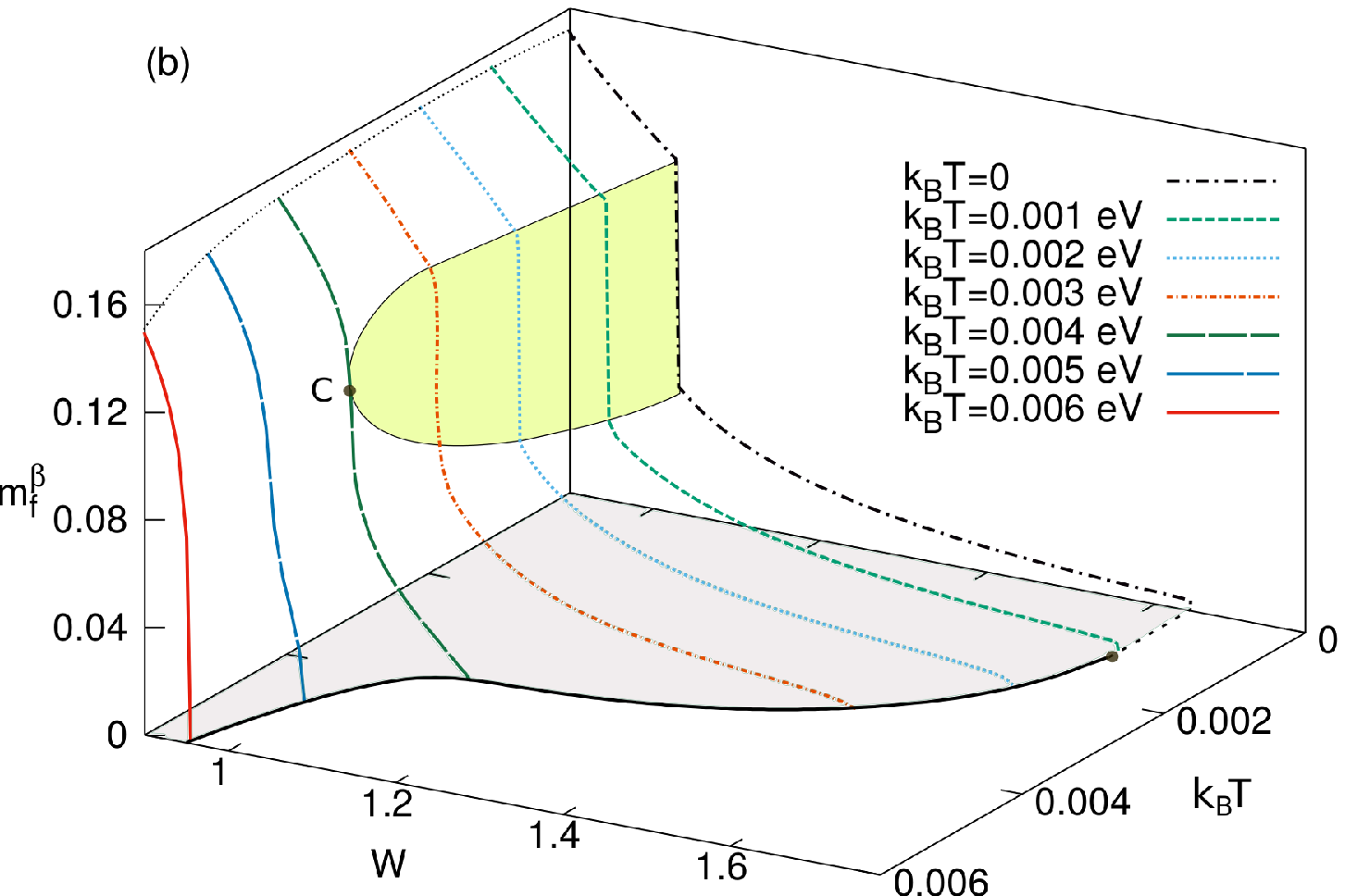}
\caption{(a) The dependence of the order parameter $m_{\alpha}$ on the band width $W$ and temperature $T$. (b) The same for the order parameter $m_{\beta}$.}
\label{fig1}
\end{figure}

The numerical calculations were performed using a $\underline{k}$-independent hybridization $V_{\alpha}(\underline{k})=V_{\alpha}$ and a total occupancy of
$N_{tot}=1.609$.  This value of the total number of electrons was chosen such that the $5f$ bands are close to half-filling so that the instability of the paramagnetic state preferentially produces N\'eel Antiferromagnetism. For wide conduction bands, due to an approximate electron-hole symmetry about f occupancies at $n_f=2$ \cite{hastatic}, similar phase diagrams are expected for $N_{tot}>2$. The parameters were chosen simply to highlight the existence of competing AF phases and their multicritical points and are not intended to be representative of any particular compound. The tight-binding parameters used are $\tilde{\epsilon}_{f}=0.3$ eV, $t_d=W_d/6$ eV, $t_f=W_d/20$ eV, $V_{\alpha}=1/10$ eV and $W_f/W_d=0.3$  where $2W_{d(f)}$ is the width of the conduction (5f) band. From here-on, we adopt the notation $W_d=W$. We also assume that the band widths $W_{d(f)}$ are sensitive to external pressure.

\begin{figure}[!t]
\centering
\includegraphics[angle=0,width=8.5cm]{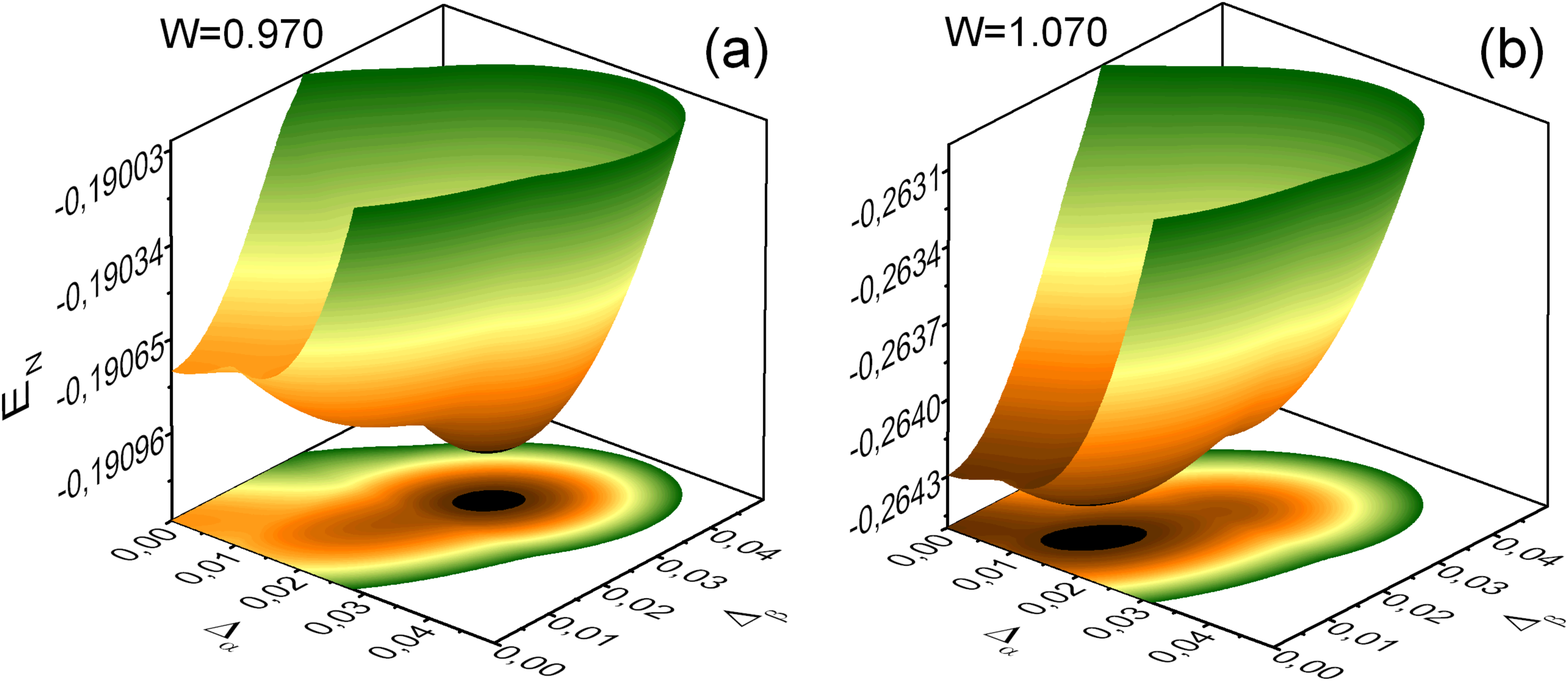}
\includegraphics[angle=0,width=8.5cm]{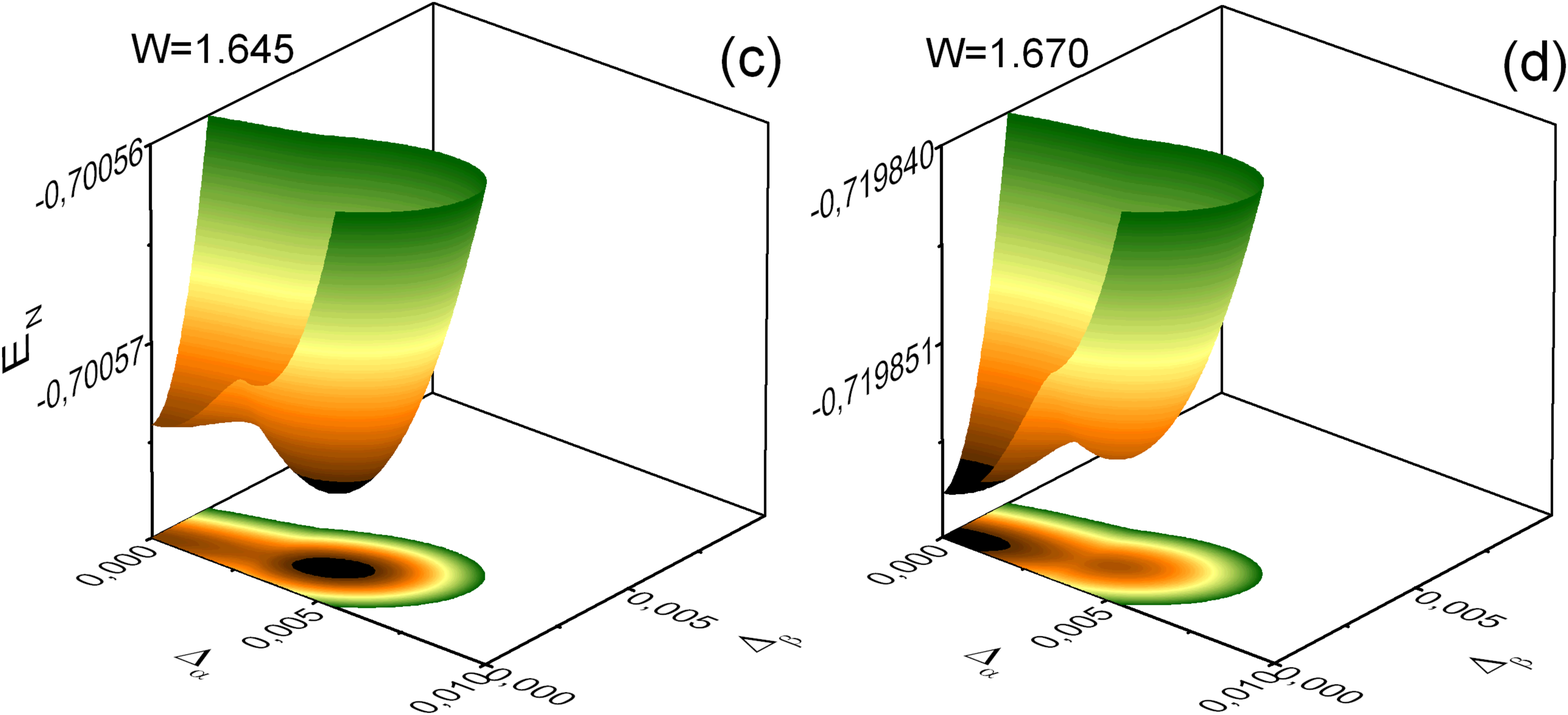}
\caption{The zero temperature value of the energy per unit cell. The projection of the ground state energy on the plan of the gaps show dark regions, in which the self-consistent solutions for the gaps $\Delta_{\alpha}$ and $\Delta_{\beta}$ are to be found. The sequence of panels shows the competition between the two types of orders as the band width $W$ is varied.}
\label{fig3}
\end{figure}

Phase diagrams are constructed from the self-consistent solutions of equations (\ref{mb}) and (\ref{ma}) for the order parameters $m^{\chi}$. The effect of Hund's Rule exchange interaction strength $J$ on the boundaries of the phases AF$_1$, AF$_2$ and PM, as $W$ increases, is shown in figure \ref{fig01}, for $T=0$. The phase AF$_1$ is characterized by $m^{\beta}>m^{\alpha}$ while the AF$_2$ denotes the phase where $m^{\alpha}>m^{\beta}$. The first-order line AF$_1$ $\rightarrow$ AF$_2$ ends at a quantum triple point located  at $(J/U)_{tri}\approx 0.07$ and $W_{tri} \approx 0.85$ where the AF$_1$, AF$_2$ and PM phases coexist.  We shall now remark on the effect that Hund's rule exchange has on the phase diagram. For $J=0$, $m_{\beta}$ and $m_{\alpha}$ are completely independent (see Eqs. (\ref{mb}) and (\ref{ma})). In this case,  the condition for opening of gaps is
satisfied for both bands 
in the region of $W \lesssim 0.8$, but for $W\gtrsim 1.1$ only the $\alpha$ band satisfies such condition.
That is, in the AF$_2$ phase the order parameter only has $\alpha$ character, i.e., $m_{\alpha}>m_{\beta}=0$. In this case, the transitions are  AF$_1$ $\rightarrow$ PM and  PM  $\rightarrow$ AF$_2$, as the band width $W$ is increased. 
For $J$  finite but small,  $m_{\alpha}$ and $m_{\beta}$ are both finite in the AF$_2$ phase. However,  the conditions for opening the gap
for both bands are still not satisfied within the intermediate interval of $W$. As $J$ further increases, the coupling between the two order parameters also increases. Above a certain value of $J$, 
the condition for opening of gaps
for both $\alpha$ and $\beta$ bands are fully recovered. This shows that there is a threshold of $J$, above which the direct transition AF$_1$ $\rightarrow$ AF$_2$ occurs. The region above the threshold is the focus of the present investigation. Therefore, from now on, we use $J=U/5$ with $U=0.165$ eV.

The staggered magnetizations $m^{\alpha}$ and $m^{\beta}$ at finite temperatures are shown in figures \ref{fig1}(a) and \ref{fig1}(b), respectively. For $k_BT=0$, both order parameters exhibit two discontinuities, one at $W\approx 1.05$ and another at $W\approx 1.65$. These discontinuities indicate the occurrence of first-order phase transitions. The first transition, occurring at $W\approx 1.05$, is between two types of antiferromagnetic phases. It should be remarked that $m_{\alpha}\lessgtr m_{\beta}$ implies that $\Delta_{\alpha}\lessgtr \Delta_{\beta}$. There is another phase transition AF$_2$ $\rightarrow$ PM. The inset in figure \ref{fig1}(a) exhibits, in detail, the region of $W$ and $T$ where this last transition takes place.

In Fig. \ref{fig3}, the dependence of the $T=0$ energy per unit cell $E_N$ on the possible values of the gaps is shown for particular values of $W$. The ground state is determined by the global minimum. The figure illustrates the competition between the AF$_1$ ($\Delta_{\beta} > \Delta_{\alpha}$), AF$_2$ ($\Delta_{\beta} < \Delta_{\alpha}$) and PM phases. The energies $E_N$ are projected on to the $\Delta_{\alpha}\times\Delta_{\beta}$ plane so the darkest regions serve as indicators for the values of $\Delta_{\alpha}$ and $\Delta_{\beta}$ that correspond to the ground states. Figures \ref{fig3}(a) and \ref{fig3}(b), show the interchange of global minimum between AF$_1$ and AF$_2$ phases. In Figures \ref{fig3}(c) and \ref{fig3}(d), $E_N$ is shown for two different values of $W$ confirming the existence of a new first-order transition AF$_2$ $\rightarrow$ PM. For $W=1.645$, the global minimum of the energy $E_N$ correspond to non-trivial solution while for $W=1.67$ the global minimum is found for the trivial solution.

\begin{figure}[t]
\centering
\includegraphics[angle=0,width=8.5cm]{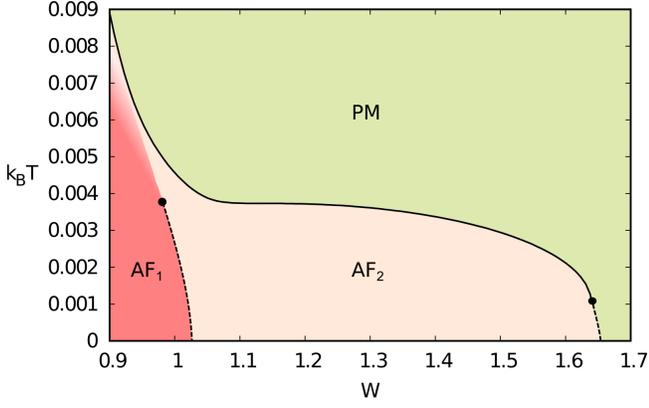}
\caption{The phase diagram for the temperature versus the band width $W$. The solid and the dashed lines denote second-order and first-order transition, respectively.
}
\label{fig2}
\end{figure}
\begin{figure}[th]
\centering
\includegraphics[angle=0,width=8.5cm]{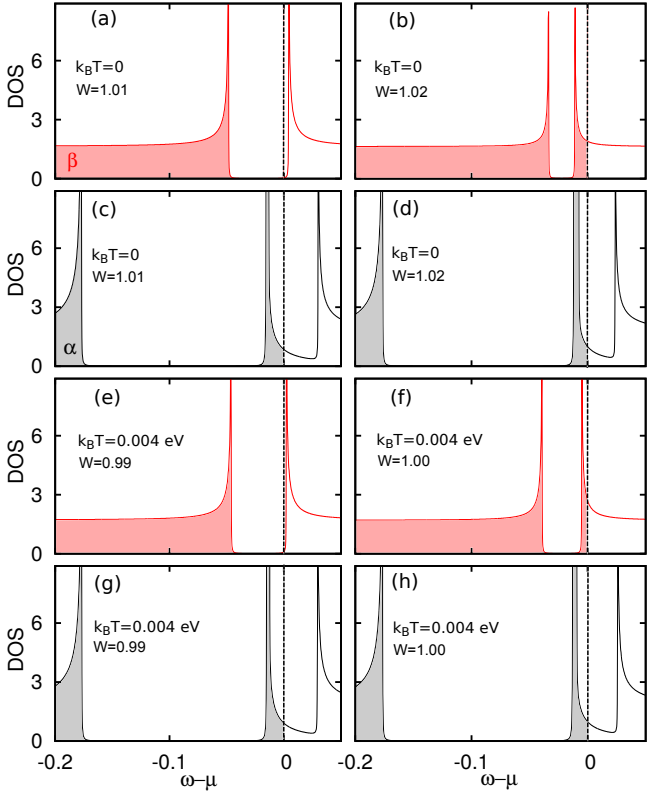}
\caption{The panels show the $\alpha$ and $\beta$ partial densities of states for values of $W$ closed to the dashed line (close AF$_1$) of the phase diagram (see figure \ref{fig2}).}
\label{fig3.3}
\end{figure}
\begin{figure}[t]
\centering
\includegraphics[angle=0,width=8.5cm]{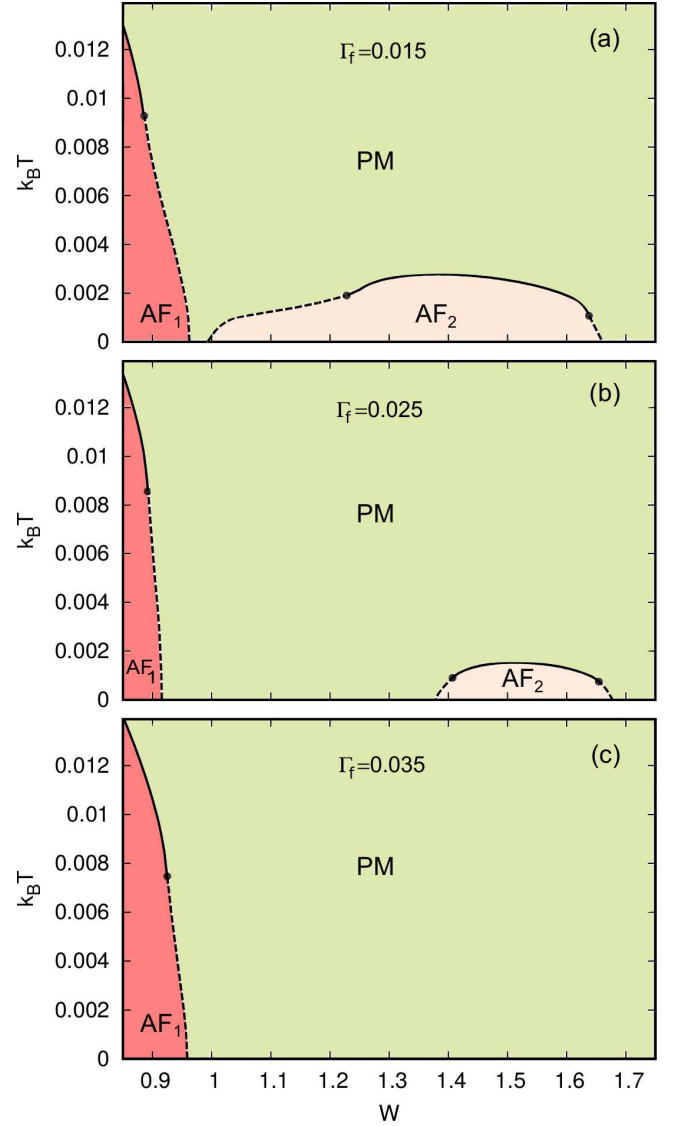}
\caption{The phase diagram of the temperature versus the band width $W$ for different values of $\Gamma_f$. The solid lines denote second-order transitions while the dashed lines denote first-order transitions. }
\label{fig5}
\end{figure}

Based on the Eq. (\ref{StoSDW}), the behaviour of the order parameters $m^{\alpha}$ and $m^{\beta}$, and the free energy given in terms of gaps $\Delta_{\alpha}$ and $\Delta_{\beta}$, it is possible to construct the phase diagram shown in figure \ref{fig2}. Firstly, there is a second-order transition at the N\'eel temperature $T_N$ (denoted by a solid line) which is marked by the formation of the AF gaps. Secondly, for $0.95<W<1.05$, there is a direct first-order transition AF$_1$ $\rightarrow$ AF$_2$ which ends at a CEP located at $k_BT_{CEP}\approx 0.0038$ and $W_{CEP}\approx 0.9826$. 
Our analysis shows that the Hund's Rule exchange interaction produces a term in the free energy which is bi-linear in $m_{\alpha}$ and $m_{\beta}$. From Eqs. (\ref{freeenergy})-(\ref{freeenergy0}), an expansion of the free energy occurs in terms of even powers of the gaps which are linear combinations of both $m^{\alpha}$ and $m^{\beta}$ (see Eq. (\ref{phi1})). We remark that in the range $T_{CEP}<T<T_N$, the jump in the order parameters becomes smooth and the two AF phases can be continuously connected by a path which bypasses the CEP. The phase diagram is completed by a third line of transitions AF$_2$ $\rightarrow$ PM which occurs for $1.6<W<1.7$. The line of transitions changes from a second-order to a first-order transition at a  tricritical point (TCP) located at $k_BT_{TCP}\approx 0.0011$ and $W_{TCP}\approx 1.6410$.

The partial Densities of States (DOS) shown in Figure \ref{fig3.3} are for band widths and temperatures in close proximity to the dashed line  which separates the phases AF$_1$ and AF$_2$ at $T=0$, and also, at temperatures where the two AFs can be connected smoothly. 
For the $\beta$ f-electron there is an insulator $\rightarrow$ metal transition at which the $\alpha$ states maintain their metallic characters. The electronic structure transition is observed to start 
 at T=0,
exactly coincides  with the line of the first-order transitions AF$_1$ $\rightarrow$ AF$_2$,  persisting above the CEP.

\begin{figure}[t]
\centering
\includegraphics[angle=0,width=8.5cm]{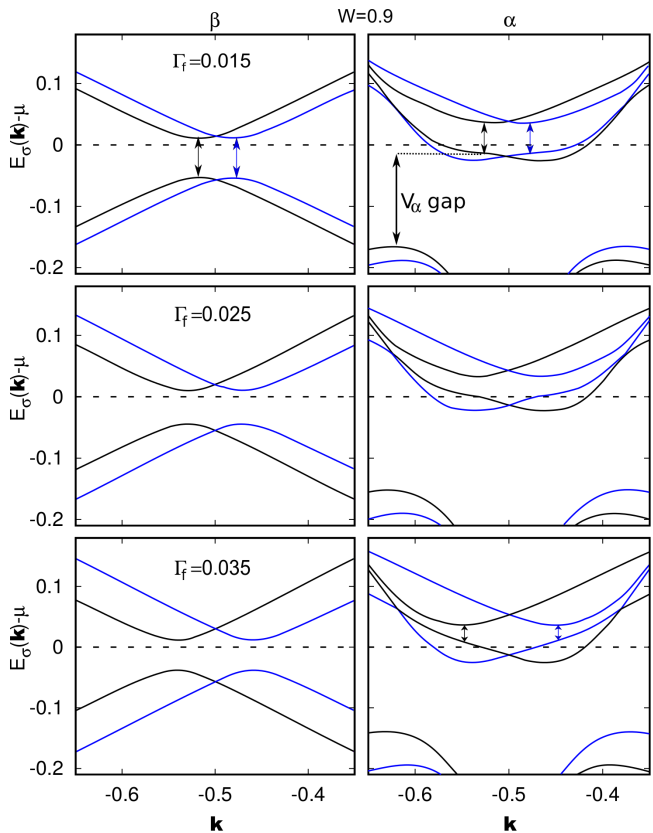}
\caption{The electronic dispersion relations near the gaps (along the direction $(\pi,\pi,\pi)-(0,0,0)$) for $W=0.9$, $T=0$ and different values of $\Gamma_f$. The blue and black colors represent the up and down spin sub-bands, respectively. The double arrows indicate where the AF gaps occur.}
\label{fig8}
\end{figure}
\begin{figure}[t]
\centering
\includegraphics[angle=0,width=8.5cm]{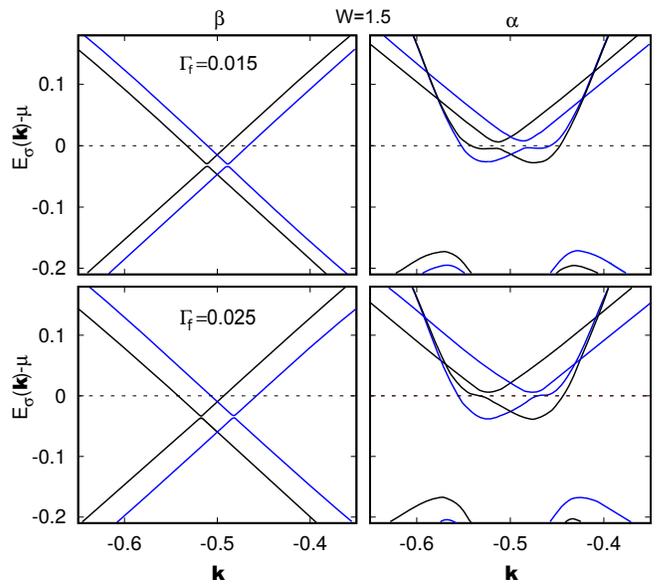}
\caption{The dispersion relations near the electronic gaps (along the direction $(\pi,\pi,\pi)-(0,0,0)$) for $W=1.5$, $T=0$ and different values of $\Gamma_f$. The blue and black colors represent the bands with spin-up and spin-down sub-bands respectively.
}
\label{fig9}
\end{figure}

The zero field magnetic phase diagram changes drastically when a transverse field $h_x$ is applied. The resulting $k_BT$ versus the band width $W$  phase diagram is shown in Figure \ref{fig5} where the values of $\Gamma_f$ are directly proportional to $h_x$ (see Eq. (\ref{Gammaf})). The main effect of $\Gamma_f$ is to separate the phases AF$_1$ and  AF$_2$ creating a dome-shaped region for this phase with two TCPs. As $\Gamma_f$ increases, the AF$_2$ domed-shaped region decreases until its complete suppression. We remark that $m^{\alpha}$ is less affected by $\Gamma_f$ in the region of $W\lesssim 1$ than in the region of $W\gtrsim 1$. For $\Gamma_f=0.035$, $m^{\alpha}$ is completely suppressed  for  $W\gtrsim 1$. In fact, the behaviour of the order parameters $m^{\alpha}$ and $m^{\beta}$ are closely related to  the condition for opening the gap
for the $\alpha$ and $\beta$ bands. The magnetic field produces a $\underline{k}$-dependent shift which depends on the spin $\sigma$. Also, the Fermi energy ($E_F$) is shifted to higher energies. Therefore, for sufficiently high values of $\Gamma_f$, the Fermi surface is no longer nested. Nevertheless, the electronic characters of the AF$_1$ and AF$_2$ phases, are unchanged.

The evolution of the phase diagram of figure \ref{fig5} can be better understood in terms of  condition for opening of gaps in 
$\beta$ and $\alpha$ bands.
For $\Gamma_f=0$, the $\beta$ sheet of the Fermi surface is nested when $E_f^{\beta}(\underline{k})=E_f^{\beta}(\underline{k}+\underline{Q})=\mu$.  The presence of $h_x$ produces a $\underline{k}$-dependent spin splitting of the dispersion relation  which
can result in a shift of $\mu$.
The evolution of the gapped regions of $\beta$ and $\alpha$ bands with increasing $\Gamma_f$, is shown in Figures \ref{fig8} and \ref{fig9}. The $\alpha$-bands involve the hybridization $V_{\alpha}$ which also affects the band's dispersion relation. In figure \ref{fig8}, the dispersion relation is calculated for a  band width of $W=0.9$, which places the system in the AF$_1$ phase (see figure \ref{fig5}). The Fermi energy $E_F$ ($\mu =E_F$) is positioned within the gap in the $\beta$-band dispersion relation for all values of $\Gamma_f$. On the other hand, at the gapped region, the extent to which the $\alpha$-band dispersion relation dips below the Fermi energy 
 decreases with increasing $\Gamma_f$. As a consequence the nesting of the $\alpha$ band is affected more strongly than the $\beta$ band. The results show that the AF phases are more stable than the paramagnetic phase, if the Fermi energy (or $\mu$) is inside  of both, or  either one or other of the $\alpha$ or $\beta$ gaps.

For  $W=1.5$ (AF$_2$ phase), we find a different situation, shown in figure \ref{fig9}. The gap in the $\beta$ band is always below the Fermi energy, whereas the Fermi energy lies within the gap of the $\alpha$ band. However, as $\Gamma_f$ increases, the Fermi energy tends to move to the bottom of the $\alpha$ gap, until for $\Gamma_f=0.035$ (not shown here), the Fermi energy falls below the gap as in the $\beta$ band case. When both gaps are below the Fermi energy the bands are not nested, and the paramagnetic phase is more stable.

\section {Conclusions}
\label{6}

In this work, we have investigated the unfolding of an itinerant antiferromagnetic phase and the subsequent emergence of multicritical points due to the competition between the 
unfolded phases that appear in a 
model suitable to describe uranium compounds. 
 This model
describes 
two  narrow
$5f$ 
bands ($\alpha$ and $\beta$) hybridized asymmetrically with a single conduction band. Besides the direct Coulomb\ interaction between electrons in the same 
$5f$ 
band, there is a Hund's rule exchange interaction between electrons in the different ones.
As main results, 
we have obtained 
temperature vs. pressure phase diagrams with and without the application of a magnetic field. 
We have assumed that pressure is associated with bandwidth variation. Moreover, since the order parameter has an Ising-like anisotropy, the magnetic field is considered transversal to this anisotropy direction.

 Therefore, we show that there is the unfolding of the antiferromagnetic phase into two distinct AF$ _1 $ and AF$ _2 $ phases. These phases are characterized by having finite staggered magnetization for both bands. Thus, AF$_1$ and AF$_2$ are given by $m_{\beta}>m_{\alpha}$ and $m_{\alpha}>m_{\beta}$, respectively.  The role of the Hund's rule exchange interaction is essential in producing these two types of phases since this interaction couples the staggered magnetization in different bands. 
Most important for the purpose of this work is that from this kind of phase competition emerges a particular set of multicritical points.
In absence of field there is a critical end point (CEP) and a tricritical point (TCP), respectively, in the phase transitions AF$_1$ $\rightarrow$ AF$_2$ and AF$_2$ $\rightarrow$ PM  as the bandwidth increases. The presence of the CEP in our phase diagram is in accordance with the description based on the generic two order parameter Landau free energy described in Ref. \cite{Mineev2005} where the order parameters were assumed to be odd under time-reversal. 
 For finite magnetic transverse field $h_x$, 
the phase diagram is drastically changed. The direct transition between the AF phases is replaced by a re-entrant sequence of transitions  AF$_1$ $\rightarrow$ PM $\rightarrow$ AF$_2$ $\rightarrow$ PM. 
 The AF$_2$ phase acquires a dome shape. 
 As consequence, the initial set of multicritical points is also completely changed. For instance, the CEP is suppressed.
 The AF$_1$ line transition has one TCP while the dome shaped AF$_{2}$ line transition has two TCPs. 
 All TCPs are effected relatively weakly by further increases of $h_x$. In fact, the dome is gradually suppressed by the field until its complete disappearance. 
$\beta$ (related to the respective bands).

 It should be remarked that the shape of the phase diagram with and without $h_x$ can be related to changes in the electronic structure. For $h_x=0$, as the bandwidth of two $5f$ bands increase, our results show that the $\beta$ $f$-electrons undergo a insulator $\rightarrow$ metal transition while the $\alpha$ electrons maintain their metallic characters. 
 Thus, our electronic structure can be described in terms of a transition between half to full metallicity.
 Below the CEP, this transition exactly coincides with the AF$_1$ $\rightarrow$ AF$_2$ first-order transition. 
The analysis of the electronic dispersion relation shows that the presence of a magnetic field $h_x$ does not change the  metallic character of both AF$_1$ and AF$_2$ phases, at least for the range of $h_x$ considered in the present work.  
 
To conclude, we highlight that  
our results show 
a detailed evolution of multicritical points when pressure and magnetic field are applied simultaneously. As far  we know, there are not much theoretical results in the literature showing this particular evolution. 
Although our results refer to a specific model of two $5f$ degenerate narrow bands, they can shed light on the growing field of the multicritical (classical and quantum) points in the physics of Uranium compounds.

\section*{Acknowledgments}

The present study was supported by the Brazilian agencies Conselho
Nacional de Desenvolvimento Cient\'{\i}fico e Tecnol\'ogico (CNPq), Coordena\c c\~ao de Aperfei\c coamento de Pessoal de N\'{\i}vel Superior (CAPES) and Funda\c c\~ao de Amparo \`a pesquisa do Estado do RS (FAPERGS).

\appendix
\section{}\label{appendixa}

In the case  where $V_{\beta}=0$, by using Eq.(\ref{matrix}), the Green's function can be explicitly written as

\begin{eqnarray}
 &G&_{ff,\sigma\sigma'}^{\beta\chi'}(\underline{k}, \underline{k'},\omega) =  \delta^{\beta\chi'}\delta_{\underline{k},\underline{k'}}
 \delta_{\sigma\sigma'}\times \\
 & &\frac{[D_{0-\sigma}^{\beta}(\omega, \underline{k})(\omega-\tilde{E}_{f}^{\beta}(\underline{k}+\underline{Q}))
 -\Gamma_{f}^{2}(\omega-\tilde{E}_{f}^{\beta}(\underline{k}))]}{|A^{\beta}|}
 \nonumber,
\label{A1}
\end{eqnarray}
 \begin{eqnarray}
  G_{ff,\sigma\sigma'}^{\beta\chi'}(\underline{k}+\underline{Q},\underline{k'},\omega)& = & \delta^{\beta\chi'}
 \delta_{\underline{k}+\underline{Q},\underline{k'}}
 \delta_{\sigma\sigma'}\times\nonumber\\
& & \frac{[\Gamma_{f}^{2}\phi^{\beta}_{-\sigma}+\phi^{\beta}_{\sigma} D_{0-\sigma}^{\beta}(\omega, \underline{k})]}{|A^{\beta}|}
\nonumber\\
\label{A2}
 \end{eqnarray}
where
\begin{eqnarray}
D^{\beta}_{0\sigma}(\omega, \underline{k}) =(\omega-\tilde{E}_{f}^{\beta}(\underline{k}+\underline{Q}))(\omega-\tilde{E}_{f}^{\beta}(\underline{k}))-
 (\phi^{\beta}_{\sigma})^{2}\nonumber \\
\label{A3}
\end{eqnarray}
and
\begin{eqnarray}
 |A^{\beta}|&=& D_{0\sigma}^{\beta}(\omega, \underline{k})D_{0-\sigma}^{\beta}(\omega, \underline{k}) +\Gamma_{f}^{2}[\Gamma_{f}^{2}-2\phi^{\beta}_{\sigma}\phi^{\beta}_{-\sigma}]\;\;\nonumber\\ & &
 -\Gamma_{f}^{2}[(\omega-\tilde{E}_{f}^{\beta}(\underline{k}))^{2}
 +(\omega-\tilde{E}_{f}^{\beta}(\underline{k}+\underline{Q}))^{2}].
\label{A4}
\end{eqnarray}

Moreover
\begin{eqnarray}
 G_{ff,\sigma\sigma'}^{\alpha\chi'}(\underline{k},\underline{k'},\omega) = \delta^{\alpha\chi'}\delta_{\underline{k},\underline{k'}} &\delta&_{\sigma\sigma'}
 \frac{(\omega-\epsilon(\underline{k}))}{|A^{\alpha}|}
 A_{1\sigma}^{\alpha}(\omega, \underline{k}),\nonumber\\
\label{A5}
\end{eqnarray}
\begin{eqnarray}
 G_{ff,\sigma\sigma'}^{\alpha\chi'}(\underline{k}+\underline{Q},&\underline{k'}&,\omega) =  \delta^{\alpha\chi'}\delta_{\underline{k}+\underline{Q},\underline{k'}} \delta_{\sigma\sigma'}A_{2\sigma}^{\alpha}(\omega, \underline{k})\nonumber\\ & & \times\frac{(\omega-\epsilon(\underline{k}))(\omega-\epsilon(\underline{k}+\underline{Q}))}{|A^{\alpha}|}
   \label{A6}
 \end{eqnarray}
where
\begin{eqnarray}
A_{1\sigma}^{\alpha}(&\omega&, \underline{k})=
 D_{0}^{\alpha}(\omega, \underline{k})
(D_{0}^{\alpha}(\omega, \underline{k}+\underline{Q}))^{2}- (\gamma_{\Gamma}^{\alpha\alpha}(\underline{k}+\underline{Q}))^{2}\nonumber\\ & & \nonumber\times D_{0}^{\alpha}(\omega, \underline{k})(\omega-\epsilon(\underline{k}+\underline{Q}))^{2}
-(\phi^{\alpha}_{-\sigma})^{2}D_{0}^{\alpha}(\omega, \underline{k}+\underline{Q})\nonumber\\ & &\times(\omega-\epsilon(\underline{k}+\underline{Q}))
 (\omega-\varepsilon(\underline{k})) \nonumber\\
\label{A7}
\end{eqnarray}
and
\begin{eqnarray}
&&A_{2\sigma}^{\alpha}(\omega, \underline{k})=
\phi^{\alpha}_{\sigma}[ D_{0}^{\alpha}(\omega, \underline{k}) \
D_{0}^{\alpha}(\omega, \underline{k}+\underline{Q})-(\phi^{\alpha}_{-\sigma})^{2}(\omega-\epsilon(\underline{k}))\nonumber\\ & &\nonumber
 \times(\omega-\varepsilon(\underline{k}+\underline{Q}))]+\phi^{\alpha}_{-\sigma}\gamma_{\Gamma}^{\alpha\alpha}(\underline{k})\gamma_{\Gamma}^{\alpha\alpha}(\underline{k}+\underline{Q})(\omega-\epsilon(\underline{k}))\\ & &\times(\omega-\epsilon(\underline{k}+\underline{Q}))\nonumber\\
 \label{A8}
 \end{eqnarray}
with
\begin{widetext}

\begin{eqnarray}
|&A&^{\alpha}|= [D_{0}^{\alpha}(\omega, \underline{k}) \
D_{0}^{\alpha}(\omega, \underline{k}+\underline{Q}) -(\phi^{\alpha}_{-\sigma})^{2}(\omega-\epsilon(\underline{k}+\underline{Q}))(\omega-\epsilon(\underline{k}))]
 [D_{0}^{\alpha}(\omega, \underline{k}) \
D_{0}^{\alpha}(\omega, \underline{k}+\underline{Q}) -(\phi^{\alpha}_{\sigma})^{2}\times \nonumber\\ && \nonumber(\omega-\epsilon(\underline{k}+\underline{Q}))
 (\omega-\epsilon(\underline{k})) ] +[(\gamma_{\Gamma}^{\alpha\alpha}(\underline{k}))^{2}(\gamma_{\Gamma}^{\alpha\alpha}(\underline{k}+\underline{Q}))^{2}
 -2\phi^{\alpha}_{\sigma}\phi^{\alpha}_{-\sigma}\gamma_{\Gamma}^{\alpha\alpha}(\underline{k})\gamma_{\Gamma}^{\alpha\alpha}(\underline{k}+\underline{Q})]
  (\omega-\epsilon(\underline{k}))^{2}
 (\omega-\epsilon(\underline{k}+\underline{Q}))^{2} \\ & &
 -(\gamma_{\Gamma}^{\alpha\alpha}(\underline{k}))^{2}(D_{0}^{\alpha}(\omega, \underline{k}+\underline{Q}))^{2}(\omega-\epsilon(\underline{k}))^{2}
 -(\gamma_{\Gamma}^{\alpha\alpha}(\underline{k}+\underline{Q}))^{2}(D_{0}^{\alpha}(\omega, \underline{k}))^{2} (\omega-\epsilon(\underline{k}+\underline{Q}))^{2}\nonumber\\
 \label{A9}
 \end{eqnarray}

 \end{widetext}
where $D^{\alpha}_{0}(\omega, \underline{k})$, $\phi^{\chi}_\sigma$ and $\gamma_{\Gamma}^{\chi\chi}(\underline{k})$ are defined in Eqs. (\ref{D0}), (\ref{GapNeel}) and (\ref{gamma}), respectively. From the equations  $|A^{\beta}|=0$
and $|A^{\alpha}|=0$, the spin independent quasi-particles energies $E^{\gamma}$ where $\gamma$ is the number of solutions. For $\gamma_{\Gamma}^{\chi}(\underline{k})\approx \Gamma_f$ and $\xi_{\Gamma}^{\chi}(\underline{k})\approx \frac{ |V_{\chi}|^2}{(\omega-\epsilon_d(\underline{k}))}$, the equation $|A^{\beta}|=0$ has $\gamma=$ 1...4 while $|A^{\alpha}|=0$ has
$\gamma=$1...8.

\section{}\label{appendixa1}
The Green's function
$G_{ff,\sigma}^{\beta\beta}(\underline{k},\underline{k}+\underline{Q},\omega)$ with $h_x=0$
acquires a simple form given by :
\begin{eqnarray}
G^{\beta\beta}_{ff,\sigma}(\underline{k},\underline{k}+\underline{Q},\omega) &=& \phi^{\beta}_{\sigma} \ \bigg({ \vert
\widetilde{B}^{+}(\underline{k})
\vert^2 \over \omega \ - E_{+}(\underline{k}) } \ + \ { \vert
\widetilde{B}^{-}(\underline{k}) \vert^2 \over \omega \ -
E_{-}(\underline{k})}\bigg)\nonumber\\
\label{Gb}
\end{eqnarray}
where the spin-independent
quasi-particle bands
$E_{\pm}(\underline{k})$ are
\begin{equation}
    E_{\pm}(\underline{k})  =
    \bigg(
{\tilde{E}_{f}^{\beta}(\underline{k})+
\tilde{E}_{f}^{\beta}(\underline{k}+\underline{Q}) \over 2}\bigg)
\pm X_{\underline{k}}
\label{emm}
\end{equation}
with
\begin{equation}
X_{\underline{k}}=\sqrt{
{\tilde{E}_{f}^{\beta}(\underline{k}) -
\tilde{E}_{f}^{\beta}(\underline{k}+\underline{Q}) \over 2}
 + (U m_f^{\beta}+J m_f^{\alpha})^2}
\end{equation}
and the spectral weights $\vert \widetilde{B}^{\pm}(\underline{k})\vert^2$ in Eq. (\ref{Gb})
found  as
\begin{eqnarray}
   \vert \widetilde{B}^{\pm}(\underline{k}) \vert^2 &=& \pm {1 \over 2} \
   {1\over
\sqrt{{\tilde{E}_{f}^{\beta}(\underline{k}) -
\tilde{E}_{f}^{\beta}(\underline{k}+\underline{Q}) \over 2}
+ \ (U m_f^{\beta}+J m_f^{\alpha})^2  } }.  \nonumber \\
\label{eW}
\end{eqnarray}

On the other hand,
the $\alpha$-band has no simple form for
the Green's function $G^{\alpha\alpha}_{ff,\sigma}(\underline{k},\underline{k}+\underline{Q},\omega)$,
\begin{eqnarray}
G^{\alpha\alpha}_{ff,\sigma}(\underline{k},\underline{k}+\underline{Q},\omega)
=\phi_{\sigma}^{\alpha} \  {(\omega-\epsilon_d(\underline{k}))(\omega-\epsilon_d(\underline{k}+\underline{Q})) \over
D^{\alpha}(\omega,\underline{k})}\nonumber\\
\label{Ga}
\end{eqnarray}
with
\begin{eqnarray}
D^{\alpha}(\underline{k}, \omega) &=& D_0^{\alpha}(\omega, \underline{k}) \
D_0^{\alpha}(\omega, \underline{k}+\underline{Q})\label{Da} \\
& & -  (U m_f^{\alpha}+J m_f^{\beta})^{2}
(\omega-\epsilon_d(\underline{k}))(\omega-\epsilon_d(\underline{k} +\underline{Q}))
\nonumber
\end{eqnarray}
and
\begin{equation}
D^{\alpha}_{0}(\underline{k}, \omega)\ = (  \omega -
 \tilde{E}_{f}^{\alpha}(\underline{k}) ) \ ( \omega - \epsilon_d(\underline{k}) ) - \vert
V_{\alpha} \vert^2.
\label{D0}
\end{equation}

\end{document}